\documentclass[Universe,article,accept,moreauthors,pdftex,10pt,a4paper]{mdpi} 

\firstpage{1} 
\articlenumber{34}
\doinum{10.3390/universe3020034}
\pubvolume{3}
\pubyear{2017}
\copyrightyear{2017}
\externaleditor{Academic Editors: Mariusz P. Dąbrowski, Manuel Krämer and Vincenzo Salzano}
\history{Received: 1 February 2017; Accepted: 29 March 2017; Published: 5 April 2017}
\pdfoutput=1

\usepackage{ifpdf}
\usepackage{graphicx}
\usepackage{amsmath}
\usepackage{amsfonts}
\usepackage{amssymb}
\usepackage{array}
\usepackage{supertabular}
\usepackage{url} 
\usepackage[figuresright]{rotating}
\usepackage{array}
\usepackage{afterpage}
\usepackage{lastpage}
\usepackage{hyperref}
\usepackage{tikz}
\usetikzlibrary{calc,arrows,decorations.pathmorphing,intersections}
\usepackage{sansmath}
\usepackage{multirow}

 \theoremstyle{mdpi}
 \newcounter{thm}
 \newcounter{ex}
 \newcounter{re}
\usepackage{microtype}
\usepackage{booktabs}
\usepackage{multirow}
\usepackage{soul}
\Title{Evaluating the New Automatic Method for the~Analysis of Absorption Spectra Using Synthetic~Spectra}

\Author{{ Matthew~B.~Bainbridge} ${^{1,}}$* and  John~K.~Webb $^2$}

\AuthorNames{Matthew~B.~Bainbridge and John~K.~Webb}

\address{%
$^{1}$ \quad Department of Physics and Astronomy, University of Leicester, 
University Road, Leicester LE1 7RH, UK\\
$^{2}$ \quad School of Physics, University of New South Wales, Sydney, 
NSW 2052, Australia; jkw@phys.unsw.edu.au\\
}

\corres{Correspondence: mbb8@le.ac.uk}

\abstract{
\textls[-15]{We recently presented a new ``artificial intelligence'' method for the 
analysis of high-resolution} absorption spectra (Bainbridge and Webb, {\em Mon. Not. R. Astron. Soc.} {\bf 2017}, {\em 468},~1639--1670). 
This new method unifies three established numerical methods: a genetic 
algorithm ({\sc GVPFIT}); non-linear least-squares optimisation with parameter 
constraints ({\sc VPFIT}); and Bayesian Model Averaging (BMA). 
In this work, we investigate the performance of {\sc GVPFIT} and BMA 
over a broad range of velocity structures using synthetic spectra. 
We found that this new method recovers the velocity structures of the 
absorption systems and accurately estimates variation in the fine structure constant. 
Studies such as this one are required to evaluate this new method 
before it can be applied to the analysis of large sets of absorption spectra. 
This is the first time that a sample of synthetic spectra has been utilised 
to investigate the analysis of absorption spectra. 
Probing the variation of nature's fundamental constants (such as the fine 
structure constant), through the analysis of absorption spectra, 
is one of the most direct ways of testing the universality of physical laws. 
This ``artificial intelligence'' method provides a way to avoid the main 
limiting factor, i.e., human interaction, in the analysis of absorption spectra. 
}


\begin{document}



\section{Introduction}

Probing the variation of nature's fundamental constants (such as the fine 
structure constant, $\alpha$), through the analysis of absorption spectra, 
is one of the most direct ways of testing the universality of physical laws. 
Interactive methods for analysing high-resolution quasar spectra of heavy 
element absorption systems are complex and require considerable expertise. 
Recently, we presented a new ``artificial intelligence'' method for 
the analysis of high-resolution absorption spectra 
\citep{bainbridge2017}. 
Our new method unifies three established numerical methods: a genetic 
algorithm ({\sc GVPFIT}); non-linear least-squares optimisation with 
parameter constraints ({\sc VPFIT}); and Bayesian Model Averaging 
(BMA). 

This method requires evaluation before being applied to the analysis of large sets of absorption spectra. 
In particular, it is unknown how the accuracy of {\sc GVPFIT} and BMA 
is effected by the complexity of an absorption systems velocity structure. 
We investigate the performance of {\sc GVPFIT} and BMA over a broad range 
of velocity structure complexities using synthetic spectra. 
This is the first time a sample of synthetic spectra has been used to investigate 
how we analyse quasar absorption spectra. 
{Using synthetic spectra, we can provide stringent tests of the modelling process. 
When~analysing spectral data, one cannot uniquely determine the velocity structure of 
the absorbing cloud and the physical parameters are unknown. 
In contrast, with synthetic spectra, the underlying (real) velocity structure and input 
parameters are uniquely determined. 
By directly comparing our models, parameter estimates, and statistical uncertainties 
with the underlying (real) velocity structures and input values, we can establish the 
stability, precision and accuracy of our approach over a broad range of complexity 
levels in the velocity structure. 
}
Such an investigation was previously infeasible due to the time-consuming nature 
of the interactive method of absorption spectra analysis.

\section{Method}

We previously applied {\sc GVPFIT} and BMA to the analysis of a 
high signal-to-noise, high spectral resolution and complex 
absorption system at $z_{abs} = 1.839$ towards J110325-264515 \citep{bainbridge2017}. 
In this analysis, {\sc GVPFIT} was iterated for over 80 generations and 
generated a large database of candidate models over a broad range of model 
complexity. 
From this large database, we selected 37 models each corresponding to the minimum-$AICc$ model with 1 through 37 velocity components, 
{ where $AICc$ is the Akaike Information Criteria corrected for small sample size (see \citep{akaike1973,hurvich1989} and Equation (\ref{eq2})).} 
We go up to 37 because a 37-component model corresponded to the minimum-$AICc$ model for the real data. 

For each of these 37 models, we utilised the Voigt profile parameters to generate 
a synthetic spectra, with $\Delta\alpha/\alpha$ set to zero. 
The appropriate {\sc VPFIT}\footnote{R. F. Carswell and J. K. Webb, 2015, 
\url{http://www.ast.cam.ac.uk/~rfc/vpfit.html}.} output was applied to 
generate the synthetic models, and hence we convolved the synthetic spectra using 
the same instrumental profile as the real spectra. 
Using the actual error arrays from the real spectra, we assigned a Gaussian 
standard deviation to each pixel and used the Box--Muller transform approach 
\citep{box1958} to add noise to the synthetic spectra. 
The~real spectra have multiple observations at different epochs and instrumental 
settings (see~\citep{bainbridge2017}~{Table~2}); 
we~generated synthetic spectra corresponding to all of these for each of the 
selected 37 models. 
Thus, the synthetic spectra emulate the characteristics of the real 
spectra of this absorption~system. 

We then treated the synthetic spectra described above as if they were real spectra 
as described in~\citep{bainbridge2017}. 
To each spectra, we applied {\sc GVPFIT} generating a large set of models to the synthetic 
spectra. 
{The synthetic spectra were both created and fitted using turbulent b-parameters and 
using the same atomic data. 
We then estimated {$\Delta\alpha/\alpha$} using BMA, with $AICc$ providing the relative 
likelihood used to weight the contribution of each model. 
In this method, $AICc$ provides a measure of the relative quality of a model, based 
on a balance of goodness-of-fit (chi-squared) against the complexity (number of 
components compared to the number of data points) of each model. 
We define $AICc$ in the normal way (\citep{akaike1973,hurvich1989})
\begin{equation}
AICc_j = \chi^2_j + 2k + \frac{2k(k+1)}{(n-k-1)}
\label{eq2}
\end{equation}
where $k$ is the number of free parameters and $n$ is the number of data points.
Statistical uncertainties are determined from the diagonal terms of the covariance 
matrix at the best-fitting solution. 
}

\section{Results}

\begin{figure}[H]
\centering
\includegraphics[scale=0.8]{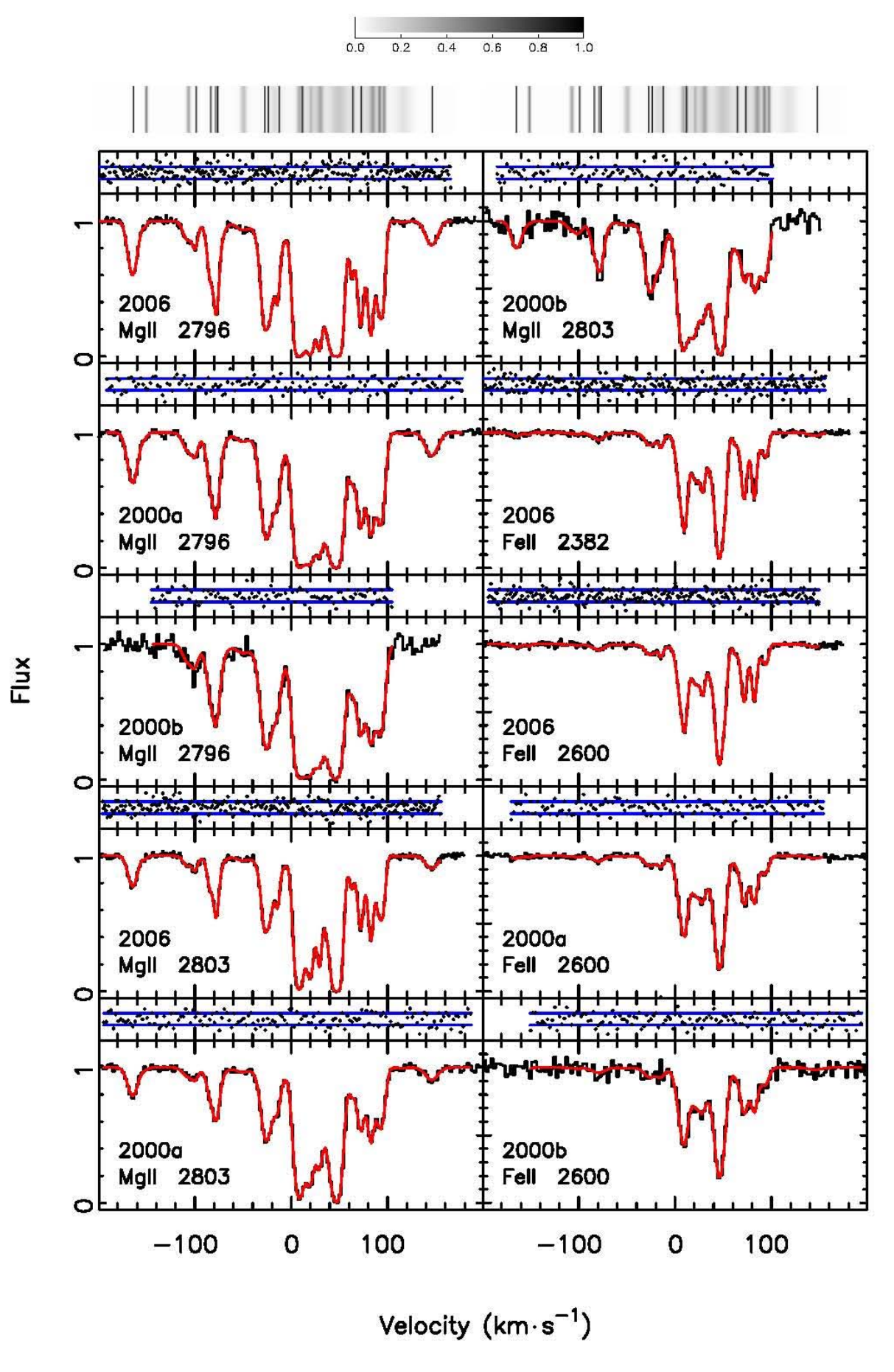}
\caption{Comparison of the { Bayesian Model Averaging (BMA)}
model and the data for the most complex synthetic 
spectrum (37~velocity components) in our sample. Above the figure is a grayscale plot showing the relative likelihood of velocity 
components at each pixel (the scale is shown at the top). 
Each row in the figure shows two panels: the top panel compares the residuals
(black points) to the $1 \sigma$ expected fluctuations (horizontal blue
lines). The bottom panel compares the BMA model (smooth red line) to the data 
(black). 
The BMA model is determined by summing over all models weighing by the relative 
likelihood of each model. 
In this figure, we show the first 10 data regions contributing to the 
$\Delta\alpha/\alpha$ analysis. 
Figure~\ref{fig2b} shows the remaining 8 data regions.}
\label{fig2a}
\end{figure}

\begin{figure}[H]
\centering
\includegraphics[scale=0.8]{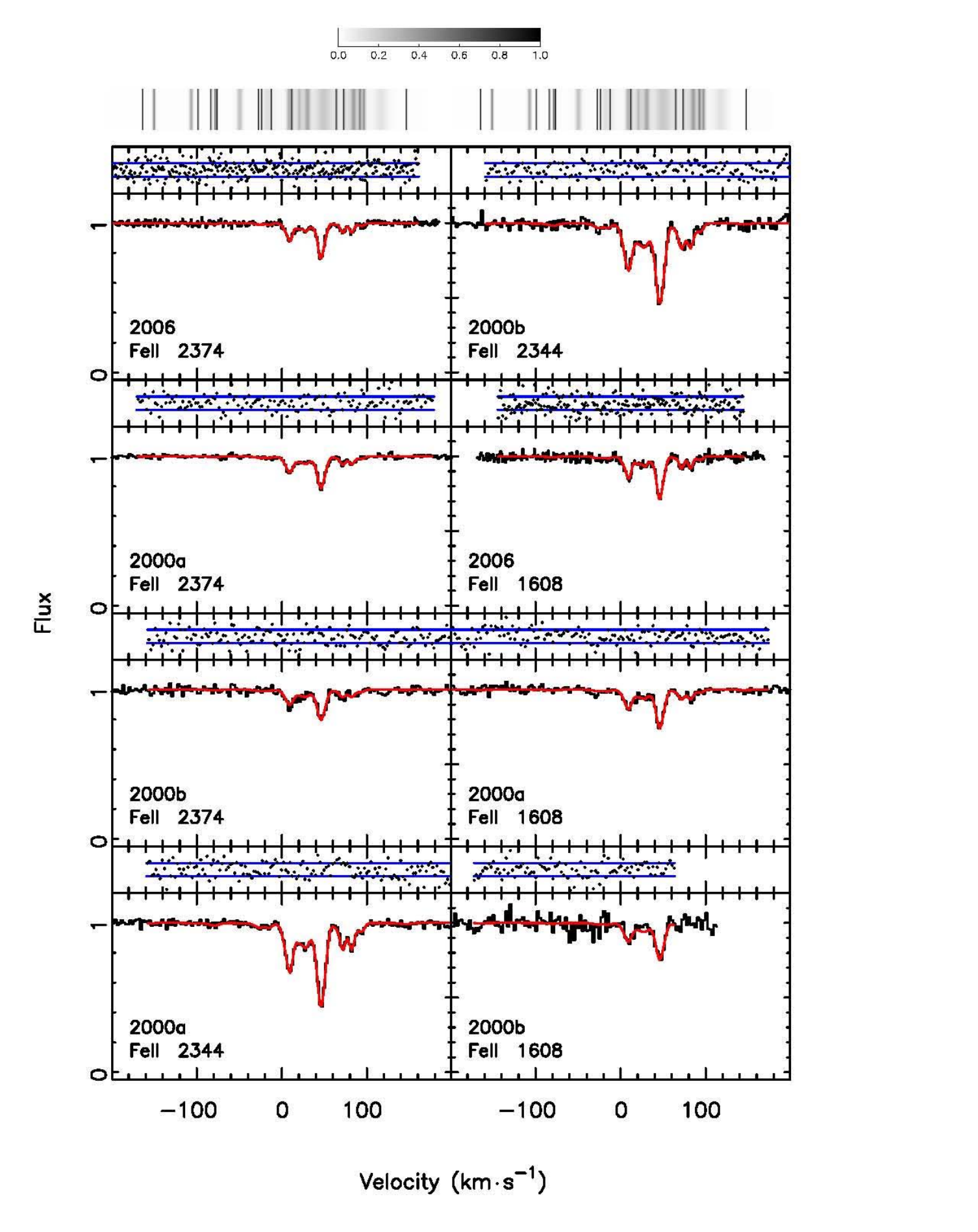}
\caption{Comparison of the BMA model and the data for the most complex synthetic 
spectrum (37~velocity components) in our sample. Above the figure is a grayscale plot showing the relative likelihood of velocity 
components at each pixel (the scale is shown at the top). 
Each row in the figure shows two panels: the top panel compares the residuals
(black points) to the $1 \sigma$ expected fluctuations (horizontal blue
lines). The bottom panel compares the BMA model (smooth red line) to the data 
(black). 
The BMA model is determined by summing over all models weighing by the relative 
likelihood of each model. 
Figure \ref{fig2a} shows the first 10 data regions contributing to the 
$\Delta\alpha/\alpha$ analysis. 
Here we show the remaining 8 data regions.}
\label{fig2b}
\end{figure}

The new ``artificial intelligence'' method, {\sc GVPFIT} and BMA, results in 
excellent fits to the synthetic spectra.  
As an example, Figures \ref{fig2a} and \ref{fig2b} illustrate the BMA model 
for the most complex synthetic spectrum we analysed, with 37 underlying (real) 
velocity components. 
These figures show the residuals are well behaved and there are no discrepancies 
between the data and the model. 
The~BMA model is determined by summing over all models for each pixel in 
the data, with the contribution of each model being weighted by its relative 
likelihood using $AICc$ (using Equations (7) and (13) from~\citep{bainbridge2017}): 
\begin{equation}
\omega(AICc_j) = \frac{\mathcal{L}(AICc_j)}{\sum_{l=1}^{S}\mathcal{L}(AICc_l)} = 
\frac{e^{-AICc_{j}/2}}{\sum_{l=1}^{S} e^{-AICc_{l}/2}}
\label{eq1}
\end{equation}
such that $\omega(AICc_j)$ is the weight of model $j$. 

Similarly, the relative likelihood of velocity components at each pixel is 
determined by summing the probability density function of each redshift parameter 
from each component in all models, weighted by relative likelihood using $AICc$ 
{ (Equation (\ref{eq1}))}. 

\begin{figure}[H]
\centering
\includegraphics[scale=0.7]{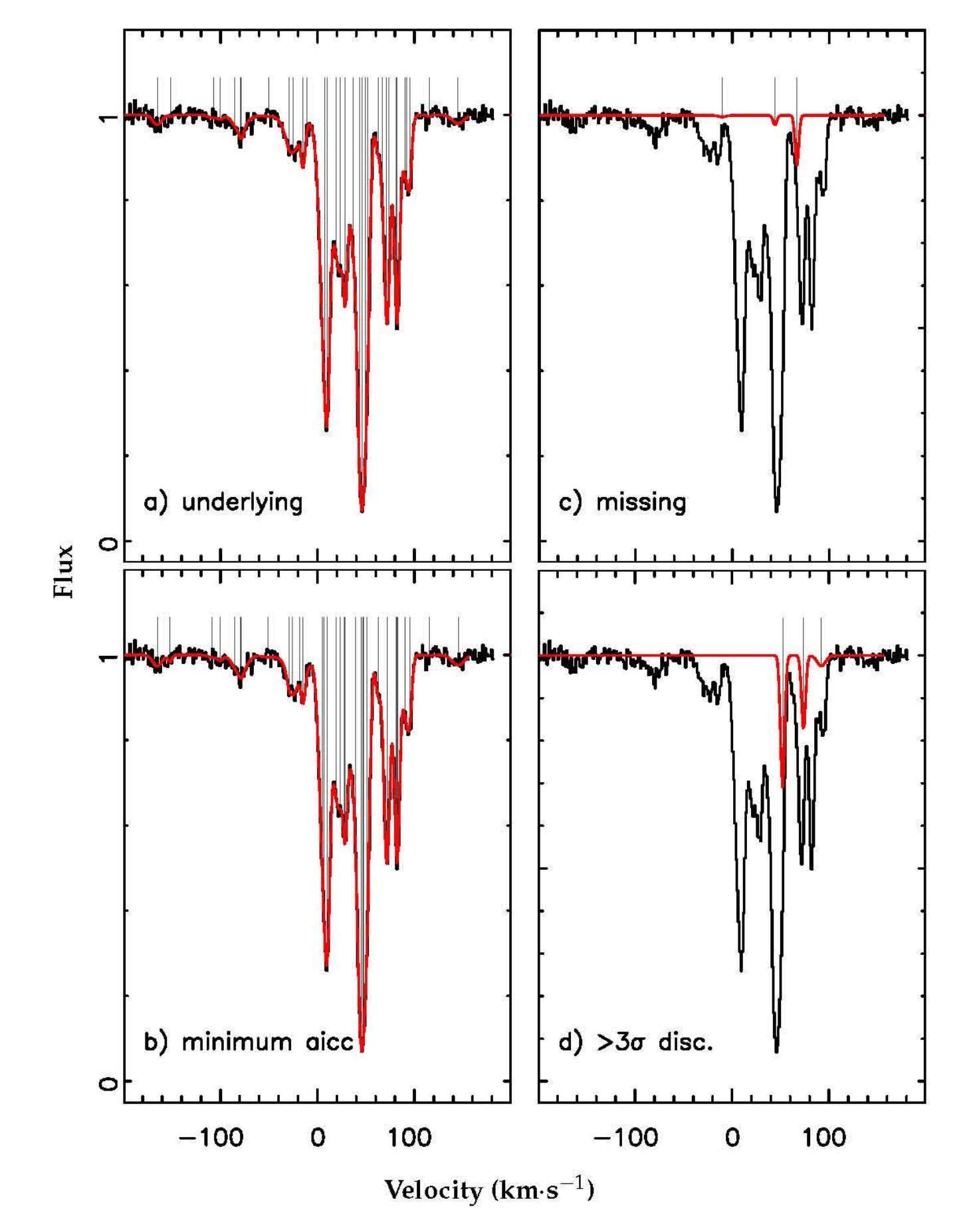}
\caption{Comparison of the minimum-$AICc$ model and underlying (real) model for the most 
complex synthetic spectrum (37 velocity components) in our sample.  The synthetic spectrum (black) shown against various models (red). 
Panel (\textbf{a}): the underlying (real) model. 
Panel (\textbf{b}): the minimum-$AICc$ model selected from the set of all models generated 
by {\sc GVPFIT}. 
Panel (\textbf{c}): the three underlying velocity components not present in the minimum-$AICc$ 
model. 
Panel (\textbf{d}): the three underlying velocity components that were inaccurately estimated 
by the minimum-$AICc$ model 
(discrepancies of >3$\sigma$ in at least one Voigt profile parameter, 
using the statistical uncertainties determined from the diagonal terms of the 
covariance matrix at the best-fitting solution). The central position of each velocity component is indicated by a tick (gray 
vertical line).}
\label{fig3}
\end{figure}

For the most complex synthetic spectra, the 37 underlying (real) velocity 
components represent a total of {148 Voigt profile parameters}, with each component contributing four Voigt profile parameters: FeII and MgII column densities, redshift and Doppler broadening $b$-parameter. 
When we compared the minimum-$AICc$ model to the underlying (real) model, we found 
that {136 parameters}, or 91.9\%, were identified. 
{\sc GVPFIT} failed to identify three velocity components, and inaccurately estimated 
{(discrepancies of >3$\sigma$ in at least one Voigt profile parameter, 
using the statistical uncertainties determined from the diagonal terms of the 
covariance matrix at the best-fitting solution) 
}
a further three velocity components. 
This is illustrated in Figure \ref{fig3}. 
The missing components are among the weakest components in the underlying model 
and are surrounded by stronger components, while the 
inaccurately estimated components are weak compared to the surrounding velocity 
components and occur in regions of dense absorption. 
This~trend is repeated throughout the entire set of 37 synthetic spectra, with 
{\sc GVPFIT} identifying 653, or 92.9\%, of the 703 underlying 
(real) velocity components. 
Additionally, four spurious (extra) weak velocity components were introduced in the 
{\sc GVPFIT} process that were not present in the original~models.

\begin{figure}[H]
\centering
\includegraphics[scale=0.7]{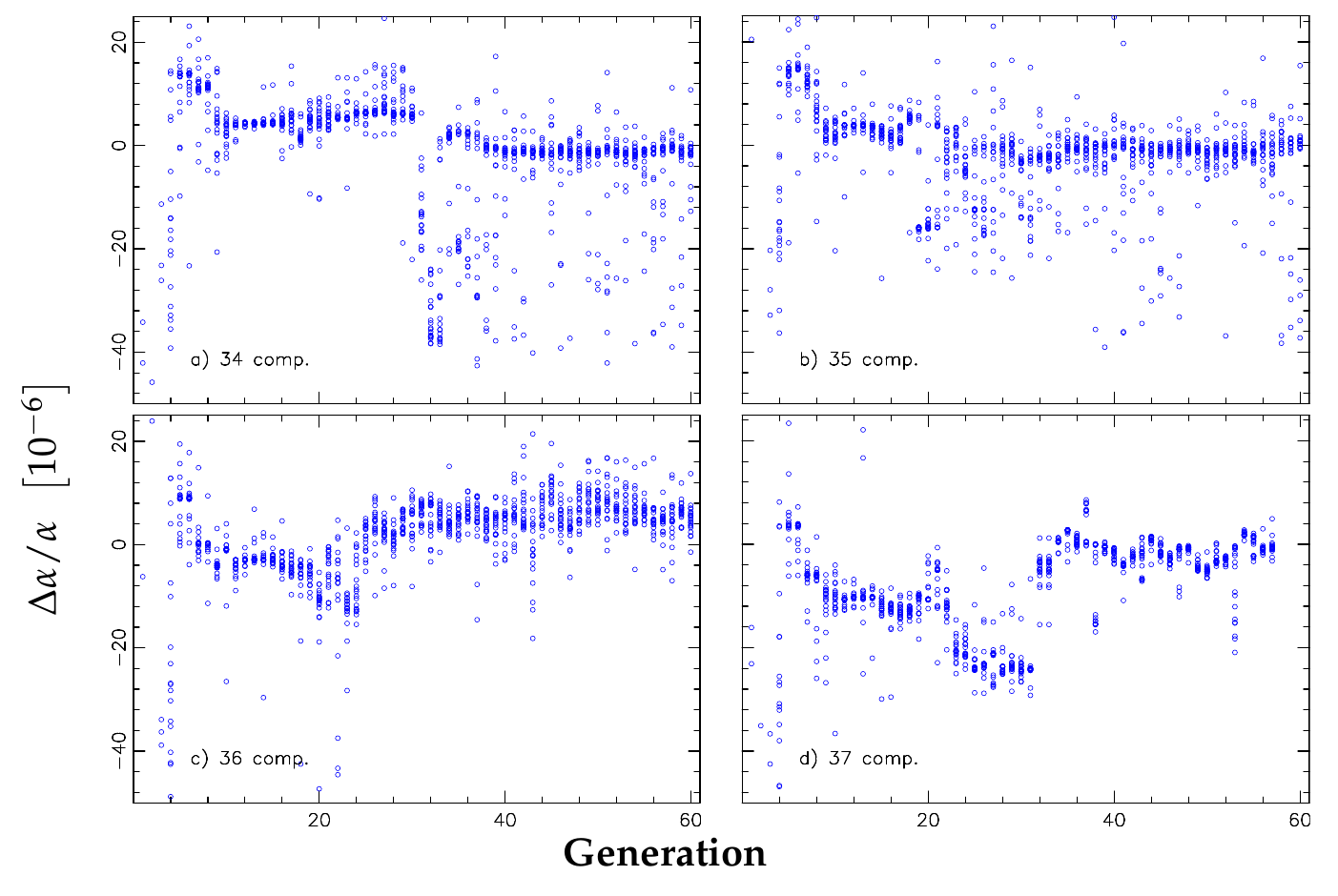}
\caption{$\Delta\alpha/\alpha$ vs generation for the four synthetic spectra with the most complex velocity structures in our sample. 
{The underlying (real) $\Delta\alpha/\alpha$ is set to zero in all cases.}
Each panel shows the $\Delta\alpha/\alpha$ estimates of the models generated 
during the analysis of each synthetic spectra. 
The underlying (real) model used to generate the synthetic spectra have 
34, 35, 36 and 37 velocity components for Panels (a), (b), (c) and (d) respectively. 
At increasing generation number, the models become increasingly complex (as 
greater numbers of velocity components are identified by {\sc GVPFIT}).}
\label{fig4}
\end{figure}

{\sc GVPFIT} recovered the underlying (real) $\Delta\alpha/\alpha$ for the synthetic 
spectra in our sample. 
Figure \ref{fig4} illustrates the $\Delta\alpha/\alpha$ estimates of all models 
generated by {\sc GVPFIT} for each of the synthetic spectra with 34, 35, 36 and 37 
underlying velocity components. 
A clear plateau is seen at $\Delta\alpha/\alpha = 0$, the~underlying (real) value. 
At lower generations, i.e., when the models are under-fit, we see conspicuous departures 
from zero. 

{Figure \ref{fig1} plots the BMA estimates of $\Delta\alpha/\alpha$ for the sample of 37 synthetic spectra.
The~inverse-variance weighted mean is $\Delta\alpha/\alpha = 0.04 \pm 0.20 \times 10^{-6}$. 
This is consistent with zero, as~expected given that the underlying (real) value of 
$\Delta\alpha/\alpha$ is zero for these synthetic spectra, and hence we found no evidence 
of a systematic bias. 
Figure \ref{fig1} also shows that the statistical uncertainties grow as the absorption 
system complexity increases, as would be expected, and this is consistent with absorption systems 
with similar quality spectral data and numbers of components from previous analyses 
\citep{webb1999,murphy2004,webb2011,king2012}. 
For~example, the statistical uncertainty from the analysis of the (real) spectral data 
for this system in King et al. (2012) \citep{king2012} is $4.0 \times 10^{-6}$ ({with} 14 velocity components and using less spectral data and different transitions) 
and in Bainbridge and Webb (2017) \citep{bainbridge2017} is $2.9 \times 10^{-6}$ 
(with 37 velocity~components). 

\begin{figure}[H]
\centering
\includegraphics[scale=0.9]{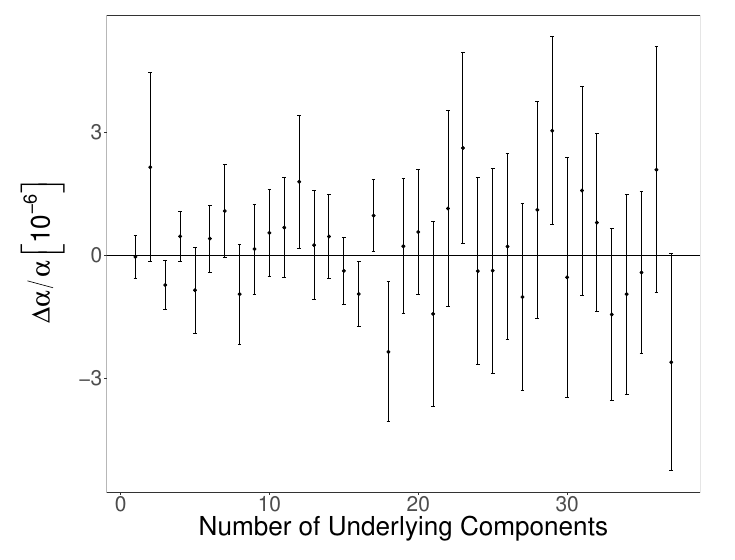}
\caption{$\Delta\alpha/\alpha$ vs number of underlying (real) velocity components for 
our sample of synthetic spectra. 
We created 37 synthetic spectra whose parameters were drawn from the 
minimum-$AICc$ models for the $z_{abs} = 1.839$ system towards J110325-264515 
(analysed in \citep{bainbridge2017}) for models that contained 1, 2, 3, ..., 37~velocity components.
$\Delta\alpha/\alpha$ is set to zero in these simulations. 
We applied Bayesian Model Averaging, BMA, to the large set of models 
generated by {\sc GVPFIT} to estimate $\Delta\alpha/\alpha$ for each of the 
37~synthetic spectra. 
The statistical uncertainties are determined from the diagonal terms of the 
covariance matrix at the best-fitting solution.}
\label{fig1}
\end{figure}

\section{Discussion}

We found that the method described in \citet{bainbridge2017}, {\sc GVPFIT} and 
BMA, recovers~the velocity structures of absorption systems and accurately 
estimates $\Delta\alpha/\alpha$ over a broad range of velocity~structures. 

{\sc GVPFIT} recovered almost all the underlying (real) Voigt profile parameters 
from the synthetic spectra (see Figure \ref{fig3}). 
The velocity components that {\sc GVPFIT} missed or inaccurately estimated are 
weak and occur in locations of dense absorption. 
We believe that it is unlikely that a human interactively fitting this set of 
synthetic spectra would perform better than {\sc GVPFIT}. 

Figure \ref{fig4} shows
interesting characteristics in the evolution of 
$\Delta\alpha/\alpha$, similar to those seen by \citet{bainbridge2017} in the 
real spectra of the $z_{abs}=1.839$ absorption system towards J110325-264515.~There appears to be an underlying linear trend in the evolution of 
$\Delta\alpha/\alpha$, with~occasional conspicuous departures 
(see Figure \ref{fig4}). These conspicuous departures exhibit a dramatic shift in $\Delta\alpha/\alpha$ 
over a small change in complexity. 
Previous interactive methods, relying on a single ``best-fit'' model, lack this 
broad picture of how $\Delta\alpha/\alpha$ evolves with velocity structure and 
may lead to a spurious estimate of $\Delta\alpha/\alpha$. 

These results also highlight the importance of having an accurate spectra error 
estimate. 
The~spectral error estimate heavily influences the statistics of the fitting 
process, as an incorrect spectral error can artificially increase or decrease 
the chi-squared ``goodness-of-fit'' statistic for a model and influence $AICc$ or 
any similar statistical criteria. 
This can lead to incorrectly estimating the number of components that are 
required to adequately fit the data and, as we have shown, have a large 
impact on the final estimate of $\Delta\alpha/\alpha$.

Future work will increase the sample size, include a more diverse set of velocity 
structures and refine the method used to generate noise for the synthetic spectra. 
The sample of synthetic spectra used in this paper is small. 
Ideally, a study of this type would consist of 1000s of synthetic spectra and the 
automated nature of the new ``artificial intelligence'' method lends itself to 
analysing large samples. 
A larger sample will allow us to increase the precision of our analysis by reducing 
the uncertainty on our weighted mean and probe for any smaller systematic bias. 
For example, a similar sample consisting of 1000 synthetic spectra should allow us 
to estimate the weighted mean below $1 \times 10^{-8}$. 
In addition, we would like to include synthetic spectra based on a broader range 
of real absorption systems, to show that this method is generalizable to a larger 
range of velocity structures, data qualities and combinations of species. 

{In addition, we expect that a more refined analysis will allow us to optimise our approach.} 
In~this work, we generate Gaussian noise using the error array from (real) spectral 
data. 
However, in~spectral data, the noise is not of Gaussian nature near zero flux and 
the noise in adjacent pixels is not independent. 
At the current level of precision with which $\alpha$ is being probed, these effects 
may become important. 

However, we believe that this work is an important contribution, giving initial 
indications that this new method is accurate and unbiased. 
The size of this sample is adequate to show there is no evidence of bias in 
$\Delta\alpha/\alpha$, when using our method, at the $2 \times 10^{-7}$ level under 
ideal circumstances (correct spectral error, high signal to noise and high resolution). 
This level of precision is two orders of magnitude smaller than the 
systematic uncertainty estimated in previous analyses of real spectral data 
(for example in \citep{king2012} $\sigma_{rand}$ is approximately $0.9 \times 10^{-5}$, 
using almost eight times the number of absorption systems). 
In addition, although GVPFIT is automated, the analysis still requires time and 
computing {resources; 
time} and resources which otherwise could be used to analyse (real) spectral data 
instead of synthetic spectra. 
Future work will extend these results, consider non-ideal circumstances and apply 
this approach to (real) spectral data.

Studies such as this one are required to test the new method of 
{\sc GVPFIT} and BMA before being applied to the analysis of large sets of 
data. 
This is the first time that synthetic spectra have been utilised to evaluate how we 
analyse absorption spectra. 
{One of the main limiting factors in the use of absorption spectra to probe fundamental 
physics is the human interaction required during the interactive modelling process. 
This human interaction involves many complex decisions, considerable expertise and 
can be very time-consuming for even a single moderately complex absorption system, 
such as a typical damped Lyman-$\alpha$ absorption system 
(such as \citep{riemer2015}). 
Furthermore, the end result can be somewhat unreliable, with the literature 
providing many examples of fits to absorption systems which are clearly inadequate. 
Much time is devoted to echelle spectroscopy of quasars on large optical telescopes and 
considerable amounts of spectra exist in telescope archives which remain unpublished, 
or which have only partially been analysed, representing a great deal of valuable 
scientific information. 
With new instruments constantly being developed, such as ESPRESSO, the quality and 
quantity of available quasar echelle spectra are only going to increase. 

Since the new method presented in {Bainbridge and Webb (2017)} \citep{bainbridge2017} 
removes the previously required human interaction, we can begin to analyse the 
ever-increasing number of quasar echelle spectra more efficiently and undertake 
projects that were previously unrealistic. 
One example of this is modelling both thermally and turbulently broadened models for 
each absorption system independently, allowing a more reliable comparison between 
models and data. The development and testing of this new ``artificial intelligence'' method 
({\sc GVPFIT} and BMA) are key to moving past the limiting factor of human 
interaction and open the way for projects that were previously unrealistic.

\vspace{6pt} 

\acknowledgments{
{This research used the ALICE High Performance Computing Facility at the University of~Leicester. 
}
}

\authorcontributions{
{ 
M.B.B. conceived, designed and performed the experiment, analyzed the data, and wrote the paper. 
J.K.W. contributed to the design of the project and the writing of the paper. 
M.B.B. and J.K.W. invented {\sc GVPFIT} and first demonstrated the application and advantages of this method.
All authors commented on the manuscript at all stages and approved the final version to be published.
}
}

\conflictsofinterest{{The authors declare no conflict of interest.}} 


\abbreviations{{The following abbreviations are used in this manuscript:
\vspace{6pt}

\noindent\hspace{-2ex}
\begin{tabular}{ll}
{\sc GVPFIT}&Genetic Voigt Profile FITting software \\
{\sc VPFIT}&Voigt Profile FITting software\\
BMA&Bayesian Model Averaging, and \\
$AICc$ &Akaike Information Criteria corrected for small sample size \end{tabular}}}

\bibliographystyle{mdpi}

\renewcommand\bibname{References}


\end{document}